\documentclass[runningheads]{llncs}
\usepackage[T1]{fontenc}
\usepackage{graphicx}
\usepackage[colorlinks=false]{hyperref}
\usepackage{booktabs}
\usepackage{multirow}
\usepackage{float}
\usepackage{array}

\begin{document}
\title{Intelligent Green Efficiency for Intrusion Detection}

%
%
\author{Pedro Pereira\inst{1}\orcidID{0009-0008-7641-1566} \and
Paulo Mendes\inst{2}\orcidID{0009-0006-6060-772X} \and
João Vitorino\inst{3}\orcidID{0000-0002-4968-3653} \and
Eva Maia \inst{4}\orcidID{0000-0002-8075-531X} \and
Isabel Praça \inst{5}\orcidID{0000-0002-2519-9859}}

\authorrunning{P. Pereira et al.}
%
\institute{Research Group on Intelligent Engineering and Computing for Advanced Innovation and Development (GECAD), School of Engineering, Polytechnic of Porto (ISEP/IPP), 4249-015 Porto, Portugal \\
\email{\{1211131,1211017,jpmvo,egm,icp\}@isep.ipp.pt}}
\maketitle              
\begin{abstract}

Artificial Intelligence (AI) has emerged in popularity recently, recording great progress in various industries. However, the environmental impact of AI is a growing concern, in terms of the energy consumption and carbon footprint of Machine Learning (ML) and Deep Learning (DL) models, making essential investigate Green AI, an attempt to reduce the climate impact of AI systems. This paper presents an assessment of different programming languages and Feature Selection (FS) methods to improve computation performance of AI focusing on Network Intrusion Detection (NID) and cyber-attack classification tasks. Experiments were conducted using five ML models - Random Forest, XGBoost, LightGBM, Multi-Layer Perceptron, and Long Short-Term Memory - implemented in four programming languages - Python, Java, R, and Rust - along with three FS methods - Information Gain, Recursive Feature Elimination, and Chi-Square. The obtained results demonstrated that FS plays an important role enhancing the computational efficiency of AI models without compromising detection accuracy, highlighting languages like Python and R, that benefit from a rich AI libraries environment. These conclusions can be useful to design efficient and sustainable AI systems that still provide a good generalization and a reliable detection.

\keywords{Green AI \and Machine Learning \and Deep Learning \and Programming Languages \and Feature Selection \and Computational Efficiency}
\end{abstract}

\section{Introduction}\label{sec:Introduction}
The rapid growth of Artificial Intelligence (AI) is creating remarkable advancements in many fields, but its impact on the environment is becoming an increasingly important issue \cite{Verdecchia_2022}. A 2020 paper by Patterson et al. \cite{Patterson2022} estimated that training a large AI model, such as OpenAI's GPT-3, could consume about 1,287 MWh of energy, leading to CO2 emissions equivalent to those produced by 125 average homes in the U.S. in one year. According to an estimate from the World Economic Forum (WEF), the information and communications technology (ICT) sector, which includes AI, by 2020 was already accountable for between 1.4\% to 5.9\% of global greenhouse gas emissions \cite{Taylor2021}. According to another estimate from the WEF, computing is expected to account for up to 8\% of global power demand by 2030 \cite{Bacchi2020}. Considering these concerns and the cruciality of intrusion detection caused by the increasing threat of cyber-attacks \cite{SecurityRisk}, this research seeks to optimize security while minimizing environmental impact \cite{EuropeanUnionAgency2}.

Green AI \cite{GreenArtificialIntelligence} highlights the importance of optimizing AI systems to minimize their energy consumption and environmental impact without compromising their performance \cite{schwartz2019greenai}. The efficient usage of computational resources can definitely lead to achieving this goal \cite{SUSTAINABLEAI}, which gets affected by various factors like the choice of a programming language \cite{EnergyProgLang}, the usage of technologies like Feature Selection (FS) \cite{FeatureSelection}, the hardware efficiency, the ML algorithm efficiency and renewable energy integration. 

FS in ML is the process of identifying and selecting the most relevant features from a dataset for model training \cite{FeatureSelection}. It can be implemented using Filter, Wrapper, and Embedded methods \cite{Lyu2023}. Filter methods evaluate features based on their statistical properties, making them efficient for high-dimensional data. Wrapper methods assess model performance with different feature sets, considering feature-model interactions, but can be computationally intensive and prone to overfitting. Embedded methods integrate FS into the model-building process, balancing accuracy and efficiency with lower computational complexity than Wrapper methods.

At the same time, the impact of FS on computational efficiency varies depending on the programming language choice. Despite being often considered, programming languages also play an essential role in impacting energy usage and computation speed \cite{EnergyProgLang}. This occurs because some languages have better optimized themselves to the hardware they are running on. Usually,compiled languages such as C++ take less time to execute than interpreted languages such as Python \cite{CompiledVSInterpreted}. Other characteristics, such as memory handling, may also affect the performance and efficiency of the algorithms.

This paper addresses the Green AI use to improve computational efficiency in Intrusion Detection Systems (IDS), seeking to optimize security while minimizing environmental impact. It assesses the choice of programming languages and FS and their relation to computational efficiency in the Green AI context and IDS. Several experiments were performed to compare quality metrics, namely accuracy, precision, recall, and F1, and footprint metrics, namely training and prediction time, in different programming languages and FS techniques.

It aims to provide useful information to researchers and practitioners to help create environmentally sustainable AI systems. This paper includes five ML models (Random Forest (RF), XGBoost (XGB), LightGBM (LGBM), Multi-Layer Perceptron (MLP), and Long Short-Term Memory (LSTM)), four programming languages (Python, Java, R, Rust), and three FS methods (Chi-Square, Information Gain (IG), Recursive Feature Elimination (RFE)).

This paper is organized into multiple sections, including details, that might help researchers to replicate this baseline, and make comparisons with their own results. \hyperref[sec:Related Work]{Section 2} presents related works. \hyperref[sec:Methods]{Section 3} details data preprocessing techniques, FS, methodology followed, and how model fine-tuning processes were carried out. \hyperref[sec:Results and Discussion]{Section 4} goes into detail regarding the results obtained on all datasets. Finally, \hyperref[sec:Conclusions]{Section 5} discusses the main conclusions that have been drawn and proposes some future research directions.

\section{Related Work}\label{sec:Related Work}
The rise of Internet of Things (IoT) has spurred significant advancements in various fields, including Network Intrusion Detection (NID). AI integration with NID has yielded positive results in enhancing security and system performance \cite{EuropeanUnionAgency}. Several papers have been conducted on the use of AI to improve IDS for IoT specifically. For instance, a systematic literature review conducted by Ferrag et al. explains the various ML algorithms used in IDS for IoT under three broad categories of supervised, unsupervised, and reinforcement learning \cite{FERRAG2020102419}. This paper especially pinpoints on the benefit of FS to improve the detection accuracy as well as the computational time of these systems.

Other studies have been made on this area, Vitorino et al. \cite{Vitorino_2022} study the impact of applying a range of the ML methods in an IoT context targeting the intrusion detection in the IoT-23 dataset, with LGBM providing better accuracy than other candidates for binary and multi-class representations. It also showed that even the completely unsupervised methods, for example Isolation Forest (iForest), worked very well in detecting previously unobserved attack. Taken from that, Silva et al. \cite{silva2024efficientnetworktrafficfeature} conducted an experiment to assess FS methods such as IG and RFE, the authors’ understanding is that the results are satisfactory because using a smaller number of features preserves performance while shortening training time. More recently, Vitorino et al. \cite{Vitorino2024} went a step further and studied adversarial robustness in IoT IDS proposing that through adversarial learning techniques can enhance system reliability against evasion attacks.

The majority of previous studies focus on AI models' detection capabilities but often overlook efficiency, a key aspect introduced by Green AI. The concept of Green AI, which emphasizes the development of energy-efficient and environmentally friendly AI models, has gained traction in recent years. Schwartz et al. note that when thinking about the intelligent systems it is also important to think of the performance and energy consumption with the future generation AI solutions suggesting factors such as carbon footprint to take in account when an model evaluation is done \cite{schwartz2019greenai}. This perspective is especially dear to IoT environments since resources are often a major limitation there.

In this context, the choice of programming languages and FS techniques plays a crucial role in the development of efficient IDS. In a research by Pereira et al. the 27 programming languages were compared regarding the runtime, memory consumption, and energy usage using ten problems from the Computer Language Benchmark Game \cite{EnergyProgLang}. The main aim of this repository is not the creation of new benchmarks, but rather, the evaluation of the performance of programming languages for the pre-specified tasks. According to research, C appeared to be the most efficient language in terms of runtime, memory usage, and power efficiency.

Another key factor that affects the efficiency of AI-based IDS is FS. Dimensional reduction of input data boosts the performance and energy efficiency of ML models. Ahmadi et al., for instance did establish that applying FS on IDS enhanced detection and reduced computational costs conducting a study using a combination of Correlation-based, IG, and Chi-square to identify the most relevant features, especially when using decision tree classifiers \cite{Ahmadi2019Efficient,Barot2014Feature,Sheen2008Network}.

Looking at IDS, specially designed for IoT, numerous works has been dedicated to detecting various types of attacks. Many data sets have been proposed and used to evaluate IDS AI models for IoT setting and among those BotIoT is highly recognized \cite{IotDiscuss}. Developed by Koroniotis et al. , the BotIoT dataset imitates several IoT-based attacks and is widely cited in the literature of the IoT field \cite{BotIotReview}. In the same way, despite the fact that the Hikari-22 is not an IoT dedicated dataset, it is a large collection of traffic data that can be very helpful to create a general benchmark to develop and evaluate IDS models. Hikari-22 is also used in IDS research community because of the variety of traffic and real life attack scenarios wherein the effectiveness of IDS in various network environments can be tested \cite{Hikari22}.

The incorporation of Green AI principles for IDS can be seen as a way of improving not just the security in networks, but efficiency as well. This way, the resource consumption of an IDS is reduced and the programming languages are chosen more appropriate by the researchers, as well as the FS techniques. Moreover, collections such as BotIoT or Hikari-22 are valuable to determine the strengths and weakness of these systems, and thus, contribute to enhance all the features, including performance and sustainability among others. 

In most cases, there is more focus on quality metrics than footprint metrics. These gaps are addressed in this paper in the sense that not only the effects of different programming languages on IDS performance are investigated under the Green AI environment but also by integrating footprint measurements into the evaluation process. Thus, our work offers a better insight into improving the performance and environmental aspects of IDS, which is immensely important for the continuous growth of Green AI in cyber security.

\section{Methods}\label{sec:Methods}
This section describes the utilized programming languages, models, FS techniques, datasets, the employed data preprocessing steps, and the evaluation metrics considered. This paper was carried out on a single machine, used to optimize and train all the models, ensuring a consistent environment during the tests. The machine has 32 GB of RAM and an 8-core CPU.

This comparative evaluation was carried out on four different programming languages to retrieve insights on the impact language choice has on the efficiency and performance of the ML models.

\subsection{Feature Selection Methods}

FS is crucial for developing efficient ML models, particularly IDS for IoT networks \cite{Rahman2021Effective,Vitorino2024Reliable}. It not only enhances performance but also reduces computational costs, aligning with the goals of Green AI. Our objective was to determine which of the features aid in the early detection of intrusions without using up a lot of resources. We employed three FS methods, Chi-Square, IG, and RFE, because they have been used in recent studies in IoT settings, obtaining good results \cite{Rihan2023Approach,Thakkar2020Attack,Yin2022IGRF-RFE:}. Preliminary experiments helped in identifying the fact that it is possible to reduce the computational load by selecting one-third of the total features present in each of the two datasets.

\textbf{Chi-Square} \cite{ChiSquare} method compares two events as to their independence – in this case, a feature and the target class. From one of the measures called chi-square, it sorts features in ascending order, aiming at those features that seem to be more independent of the class label. This technique is especially helpful when dealing with categorical data because it allows for the identification of contributing features with the largest correlations to the target variable.

\textbf{IG} \cite{IG} determine an average reduction in entropy or uncertainty of the target variable provided by the feature in question. It sorts features on the basis of their usefulness in determining the class label and makes it possible for us to deploy features that considerably enhance the model’s precision.

\textbf{RFE} \cite{RFE} is a technique of recursively eliminating the attributes that have the least weights assigned by a given model until those attributes that are deemed to be most valuable are obtained.

With these FS methods, we made certain that not only were the models efficient in terms of their performance but also resource usage, which forms a part of the attempts being made towards creating sustainable AI solution for systems security.

\subsection{Programming Languages}
Given the significant influence that programming languages can exert on the performance and efficiency of an AI model \cite{Pinto_2024}, we implemented our approach in four distinct languages: Python, Java, R, and Rust. All of these languages come with their own benefits in relation to the execution times and efficiency factors and thus are suitable for this research.

\textbf{Python} in particular, characterized by rather simple and clear syntax, has received rather significant degree of attention, particularly in the context of AI and ML. Numerous libraries like TensorFlow, and Scikit-Learn make it the most preferred language in data analysis and AI creation \cite{Raschka2020}. Python is an ideal language for data science and for creating the first versions of ML systems, computer vision programs, natural language processing tools. However, the language is interpreted and has the Global Interpreter Lock meaning it is slower when it comes to processing intensive tasks compared to compiled languages \cite{CompiledVSInterpreted}.

\textbf{R}, meanwhile, excels with it statistical computing and data analysis property and as such it is well suited for probability modeling \cite{Giorgi2022The,Paquot2020Descriptive}, simulating and data visualization which are key aspects in building AI systems. Using packages such as caret, R provides useful package even on data manipulation and visualization. It also includes packages for DL such as TensorFlow for R.

\textbf{Java} cannot be overestimated among AI agents as it gives the ability to integrate new AI components with existing systems of an enterprise using JVM without the need to rewrite existing business applications \cite{Lang2019WekaDeeplearning4j:,Vanam2021High}. Java has a number of strong libraries for instance DeepLearning4j for neural network and Weka for ML algorithms. However, it has to be understood that Java may be more verbose and as such, the process of code development may take more time than if done using languages such as Python or R.

Although relatively less popular among the AI enthusiasts, \textbf{Rust} offers performance and memory safety in high-performance computing use-cases \cite{Lin2016Rust}. But because Rust is a compiled language, it executes faster which is paramount for large models that will take ages to train and infer. It poses an ownership concept which makes it safe for memory, and it avoids crashes thus suitable for ML. However, Rust is comparatively new and has a less developed ecosystem, though it also has rather fine-grained control over what is happening in the program, and thus may be more challenging to learn.

\subsection{Datasets}
There are many datasets available to test Network Intrusion Detection Systems (NIDS) and all of them have their own properties. So for this paper we wanted to include two datasets that are comparatively recent and thematically closest, one that focuses specifically on the IoT environment and another more general dataset for NIDS research. To this end, and as we shall show in the following sections, we selected the BotIoT dataset, for IoT security studies and the Hikari-22 dataset that can be used in NIDS studies in general. This makes it possible to perform a diverse analysis of the specialized and generalized network intrusion.

\textbf{BotIoT} \cite{BotIotReview} dataset was developed as part of a project to collect and create most of the type of malicious activities that can happen in IoT network. The applications for this dataset include utilizing the Cyber Range Lab of the Australian Centre for Cyber Security, some of which are smart refrigerators, thermostats or surveillance cameras. Since the BotIoT dataset has various labeled data that can be utilized for example in ML to detect and prevent IoT based threats. The presence of different attacks and realistic IoT environments make the given dataset quite useful for IDS and other security systems investigations.

\textbf{Hikari-22} \cite{Hikari22} is the more recent than BotIoT dataset, proposed to solve the lack of publicly available up-to-date datasets for cyber security appliances. It was created from traffic captured on the internet, either as a packet or flow. Afterward, the captured traffic is compiled into a specific data type containing network-related features, including labeling. The raw data is then parsed through various pre-processing steps until it reaches its final stage. This dataset includes encrypted synthetic attacks that are lacking in many others \cite{Ferriyan2021-HIKARI}.

For both datasets, the three previously introduced FS methods were applied independently.

\subsection{Models and Fine-tuning}
Five types of ML models were chosen for this paper: RF, XGB, LGBM, MLP and LSTM. Hyperparameters were tuned for each of the models on each of datasets through a grid search which involves experimenting with the best established hyperparameter combinations. The configurations were selected based on 5-fold cross-validation and the evaluation metric was the macro-average F1-score so that the method is not sensitive to the imbalance in the sets. The following models were selected along with the fine-tuned parameters of each model:

\textbf{RF} \cite{breiman2001random} is an ensemble learning method that builds multiple decision trees during training. Each tree in the forest casts a vote, and the prevailing response is the final estimate. It reduces overfitting and increases the accuracy of classification or regression tasks. Table \ref{tab:rf_config} summarizes the fine-tuned configuration.

\begin{table}[H]
\centering
\caption{Summary of RF configuration.}
\label{tab:rf_config}
\begin{tabular}{lcc}
\toprule
Hyperparameter & BotIoT Value & Hikari-22 Value \\
\midrule
Criterion & Entropy & Gini \\
No. of estimators & 128 & 128 \\
Min samples split & 2 & 2 \\
Max. depth of a tree & None & 8 \\
Min. samples in a leaf & 1 & 1 \\
\bottomrule
\end{tabular}
\end{table}

\textbf{XGB} \cite{chen2016xgboost} is a way of implementing extreme gradient boosting techniques and is efficient and scalable. It creates models one at a time so the models it creates are corrections or improvements upon the models prior to them. Table \ref{tab:xgb_config} summarizes the fine-tuned configuration.

\begin{table}[H]
\centering
\caption{Summary of XGB configuration.}
\label{tab:xgb_config}
\begin{tabular}{lcc}
\toprule
Hyperparameter & BotIoT Value & Hikari-22 Value \\
\midrule
Booster & gbtree & gbtree \\
Learning rates & 0.4 & 0.3 \\
No. of estimators & 128 & 32 \\
Objective & multi:softmax & multi:softmax \\
Subsample & 0.8 & 0.8 \\
Tree Method & hist & hist \\
\bottomrule
\end{tabular}
\end{table}

\textbf{LGBM} \cite{ke2017lightgbm} is a gradient boosting used for developing tree-based ML algorithm. It is very efficient in performing data operations and specifically designed for handling huge data with minimum memory consumption, especially recommended for  high-dimensional. Table \ref{tab:lgbm_config} summarizes the fine-tuned configuration.

\begin{table}[H]
\centering
\caption{Summary of LGBM configuration.}
\label{tab:lgbm_config}
\begin{tabular}{lcc}
\toprule
Hyperparameter & BotIoT Value & Hikari-22 Value \\
\midrule
Learning rate & 0.1 & 0.1 \\
No. of estimators & 32 & 64 \\
Max. depth of a tree & -1 & -1 \\
Max no. of leaves per tree & 32 & 32 \\
\bottomrule
\end{tabular}
\end{table}

\textbf{MLP} \cite{almeida2020multilayer} is feedforward artificial neural networks. It also comprises more than one layer of neurons, and each neuron from a given layer is connected to all the neurons in the layer following it. They can be applied in tasks such as classifications, regression analysis, and to transform the data in to features, more especially in cases where the data is non-linear. Table \ref{tab:mlp_config} summarizes the fine-tuned configuration.

\begin{table}[H]
\centering
\caption{Summary of MLP configuration.}
\label{tab:mlp_config}
\begin{tabular}{lcc}
\toprule
Hyperparameter & BotIoT Value & Hikari-22 Value \\
\midrule
Activation & Logistic & Tanh \\
Hidden Layers & (32, 64) & (64, 32) \\
Learning Rate & 0.001 & 0.001 \\
Batch size & 64 & 64 \\
Epoch      & 200 & 200 \\
\bottomrule
\end{tabular}
\end{table}

\textbf{LSTM} \cite{graves2012long} is a kind of DL model that has the ability to remember past events, LSTM is sub type of recurrent neural network (RNN). Some of the specific features include the presence of a memory cell where data stored in the memory is held for a long duration in a retained mode of operation particularly in applications such as time series forecasting, text and language processing and analysis of sequential data. Table \ref{tab:lstm_config} summarizes the fine-tuned configuration.

\begin{table}[H]
\centering
\caption{Summary of LSTM configuration.}
\label{tab:lstm_config}
\begin{tabular}{lcc}
\toprule
Hyperparameter & BotIoT Value & Hikari-22 Value \\
\midrule
Hidden Layers & (32, 32, 32) & (16, 32, 16) \\
Dropout rate  & 0.4          & 0.5 \\
Learning rate & 0.1          & 0.1 \\
Batch size    & 32           & 32 \\
Epoch         & 128          & 64 \\
\bottomrule
\end{tabular}
\end{table}

\section{Results and Discussion}\label{sec:Results and Discussion}

This section will focus on the assessment of the implemented ML models. Here, we evaluate their performance in achieving the classification task using two main types of metrics, quality metrics - accuracy, precision, recall and F1 - and footprint metrics - training and prediction time, both in seconds. For each dataset, two tables are presented: one displaying results without FS and another showcasing the best performance achieved among the three FS models.

\subsection{BotIoT}
Table \ref{tab:BotIoT_NoFs_results} provides the obtained results for the models trained with the BotIoT dataset without FS. It is important to note that some models are not possible to implement in certain languages.

\begin{table}[h]
\caption{Obtained results for the BotIoT dataset without Feature Selection.}
\label{tab:BotIoT_NoFs_results}
\scriptsize
\setlength{\tabcolsep}{7pt}
\begin{tabular}{@{}cccccccc@{}}
\toprule
\textbf{Model} & \textbf{Lang.} & \textbf{Acc.} & \textbf{Prec.} & \textbf{Rec.} & \textbf{F1} & \textbf{Train. Time} & \textbf{Pred. Time} \\ \midrule
\multirow{4}{*}[-.7em]{RF}       & Java   & 0.9998 & 0.8906 & 0.8070 & 0.8366 & 110.234   & 5.296    \\ \cmidrule(l){2-8} 
                          & Python & 0.9998 & 0.8659 & 0.8296 & 0.8455 & 44.8593   & 1.4218   \\ \cmidrule(l){2-8} 
                          & R      & 0.9998 & 0.7687 & 0.9856 & 0.8638 & 303.7559  & 1.2278   \\ \cmidrule(l){2-8} 
                          & Rust   & 0.9998 & 0.9417 & 0.8002 & 0.8411 & 2905.2297 & 69.9858  \\ \midrule
\multirow{4}{*}{XGB}      & Java   & 0.9999 & 0.9998 & 0.9998 & 0.9998 & 1.687     & 0.047    \\ \cmidrule(l){2-8} 
                          & Python & 0.9998 & 0.8469 & 0.8174 & 0.8292 & 85.8593   & 1.625    \\ \cmidrule(l){2-8} 
                          & R      & 0.9998 & 0.8174 & 0.8485 & 0.8326 & 11.3276   & 0.1372   \\ \midrule
\multirow{4}{*}[.8em]{LigthGBM} & Python & 0.9941 & 0.5864 & 0.6978 & 0.6195 & 13.4843   & 1.5      \\ \cmidrule(l){2-8} 
                          & R      & 0.9991 & 0.719  & 0.6718 & 0.6946 & 7.3249    & 0.0991   \\ \midrule
\multirow{4}{*}[-.7em]{MLP}      & Java   & 0.9991 & 0.4988 & 0.5000 & 0.4994 & 4442.911  & 10.98    \\ \cmidrule(l){2-8} 
                          & Python & 0.9992 & 0.6351 & 0.6532 & 0.6436 & 1073.5625 & 2.125    \\ \cmidrule(l){2-8} 
                          & R      & 0.9487 & 0.4075 & 0.4857 & 0.4431 & 638.6015  & 7.826    \\ \cmidrule(l){2-8} 
                          & Rust   & 0.0115 & 0.2686 & 0.3302 & 0.2767 & 992.8198  & 140.0066 \\ \midrule
\multirow{4}{*}{LSTM}     & Java   & 0.9606 & 0.9754 & 0.4297 & 0.9061 & 444.858   & 1.485    \\ \cmidrule(l){2-8} 
                          & Python & 0.9995 & 0.7492 & 0.7492 & 0.679  & 3468.0638 & 9.8897   \\ \cmidrule(l){2-8} 
                          & R      & 0.9996 & 0.7772 & 0.9383 & 0.8502 & 34902.385 & 34.8341  \\ \bottomrule
\end{tabular}
\end{table}

The XGB and RF have the highest overall performance on quality metrics, especially in Java implementations. Nevertheless, there are certain issues related to the speed at which Java models are trained. XGB in Java presents an incredibly short train. time, while all quality metrics have values above 0.9998, beyond that LSTM in Java seems also having problems having a significantly high F1, despite its low recall. This raises questions about the reliability of Java's efficiency in training complex models with the actual state of ML libraries.

Despite these issues, is to stand up the case of Python with RF having an accuracy of 0.9998, combined with a strong F1 of 0.8455, making it a reliable model for various applications. RF in R also has an high accuracy and F1, 0.8638 which is biggest than Python, but the difference in train. time is high, 44.8593 s to Python and 303.7559 s to R. XGB despite having lightly worse results of F1 compared to RF models, in R they excel achieving a better balance between precision and result doing with a significant difference in train. time, with RF taking 303.7559 s and XGB taking only 11.3276 s. 

In terms of train. time, LGBM models take the least time in all languages, especially in R, where the model taking only 3.7039 s. However, LGBM typically doesn't provide very high accuracy and F1 compared to the other models. On the other hand, MLP models, are usually the most time-consuming, especially for models implemented in Java as the train. time is striking 4442.911 s although the model was very accurate up to 0.9991. Nevertheless, the MLP model has a considerably lower F1, noticeably lower in Java, showing the poorest performance along all indicators.

Considering the programming languages, Python proves itself reasonable and balanced language to implement ML models, providing high scores on quality metrics and, at the same time, good footprint metrics. R, thus, also get good stats, it obtained similar quality metrics compared to Python, obtaining worse footprint metrics in RF but for XGB, LGBM and MLP it obtained a better footprint metrics. Java is quite capable in certain aspects, but it is less efficient in other areas, and there is inconsistency, making it less ideal  when a steady model training is needed. Rust also showed good results in quality metrics, but footprint metrics were pretty bad.

Comparing the results of ML models with FS and ML models without FS for BotIoT it can be inferred that the general quality metric do not differ significantly across the programming languages. However, one of the most significant benefits of using FS is that it leads to reduced training and inference time, and this is observed. Table \ref{tab:BotIoT_Fs_results} provides the obtained results for the models trained with the BotIoT dataset with FS.

The patterns observed in the previous table are maintained to this one. RF and XGB keeps as the best models on quality metrics with balances footprint metrics, LGBM keeps with the best results on footprint metrics but not good results in footprint metrics and MLP with bad results in both metrics. Despite this improvements, Java continues to exhibit challenges even with FS for the XGB and LSTM models. Beyond that it is important to highlight the improvements made by FSn on Rust MLP model, without FS Rust MLP presents an F1 of 0.2767, pretty lower than the results obtained with FS, 0.7010. This global analysis highlights the necessity of the choice of FS techniques and work languages that are appropriate to the characteristics and conditions of the further application of ML models.

\begin{table}[h]
\caption{Obtained results for the BotIoT dataset with Feature Selection.}
\label{tab:BotIoT_Fs_results}
\scriptsize
\setlength{\tabcolsep}{3.6pt}
\begin{tabular}{@{}ccccccccc@{}}
\toprule
\textbf{Model} &
  \textbf{Lang.} &
  \textbf{FS Tech.} &
  \textbf{Acc.} &
  \textbf{Prec.} &
  \textbf{Rec.} &
  \textbf{F1} &
  \textbf{Train. Time} &
  \textbf{Pred. Time} \\ \midrule
\multirow{4}{*}[-.7em]{RF}       & Java   & IG         & 0.9997 & 0.8809 & 0.7936 & 0.8285 & 49.734     & 4.797    \\ \cmidrule(l){2-9} 
                          & Python & Chi-Square & 0.9998 & 0.8673 & 0.8105 & 0.8345 & 77.7812    & 1.5937   \\ \cmidrule(l){2-9} 
                          & R      & Chi-Square & 0.9998 & 0.7723 & 0.9830 & 0.8650 & 77.9851    & 0.7614   \\ \cmidrule(l){2-9} 
                          & Rust   & Chi-Square & 0.9997 & 0.8742 & 0.8140 & 0.8414 & 1697.3741  & 72.6016  \\ \midrule
\multirow{4}{*}{XGB}      & Java   & ALL        & 0.9999 & 0.9998 & 0.9998 & 0.9998 & 3.297      & 0.031    \\ \cmidrule(l){2-9} 
                          & Python & Chi-Square & 0.9997 & 0.9885 & 0.7425 & 0.7980 & 64.0468    & 1.625    \\ \cmidrule(l){2-9} 
                          & R      & Chi-Square & 0.9997 & 0.739  & 0.9883 & 0.8457 & 9.7582     & 0.1828   \\  \midrule
\multirow{4}{*}[.8em]{LigthGBM} & Python & RFE        & 0.9980 & 0.5550 & 0.6665 & 0.5819 & 4.7187     & 2.6093   \\ \cmidrule(l){2-9} 
                          & R      & IG         & 0.9995 & 0.6866 & 0.8074 & 0.7421 & 6.5932     & 0.1082   \\  \midrule
\multirow{4}{*}[-.7em]{MLP}      & Java   & IG         & 0.9991 & 0.4985 & 0.5000 & 0.4993 & 3564.804   & 8.937    \\ \cmidrule(l){2-9} 
                          & Python & RFE        & 0.9996 & 0.7466 & 0.6625 & 0.6958 & 840.0468   & 1.75     \\ \cmidrule(l){2-9} 
                          & R      & RFE        & 0.999  & 0.4998 & 0.4985 & 0.4992 & 634.985    & 7.1586   \\ \cmidrule(l){2-9} 
                          & Rust   & RFE        & 0.9995 & 0.7452 & 0.6704 & 0.7010 & 807.1418   & 110.7449 \\ \midrule
\multirow{4}{*}{LSTM}     & Java   & Chi-Square & 0.9583 & 0.9741 & 0.4255 & 0.8997 & 431.983    & 0.984    \\ \cmidrule(l){2-9} 
                          & Python & IG         & 0.9996 & 0.9544 & 0.7216 & 0.7527 & 3634.3075  & 10.2833  \\ \cmidrule(l){2-9} 
                          & R      & Chi-Square & 0.9997 & 0.7234 & 0.9896 & 0.8358 & 11286.8694 & 15.4808  \\  \bottomrule
\end{tabular}
\end{table}

\subsection{Hikari-22}
Given the issues identified with the Java XGB and LSTM models implemented in the BotIoT dataset, a subsequent analysis of these models on the Hikari-22 dataset reveals that these problems persist. Nevertheless, the XGB and RF models remain robust and prove to have the best results on the quality measures. Notably, LGBM has become an alternate model yielding nearly similar quality metrics. For instance, the F1 for LGBM, RF, and XGB are 0.9395, 0.9397 and 0.9398 respectively which are nearly equal. Table \ref{tab:Hikari-22_NoFs_results} shows the results of models trained on the Hikari-22 dataset without FS.

\begin{table}[h]
\caption{Obtained results for the Hikari-22 dataset without Feature Selection.}
\label{tab:Hikari-22_NoFs_results}
\scriptsize
\setlength{\tabcolsep}{7pt}
\begin{tabular}{@{}cccccccc@{}}
\toprule
\textbf{Model} & \textbf{Lang.} & \textbf{Acc.} & \textbf{Prec.} & \textbf{Rec.} & \textbf{F1} & \textbf{Train. Time} & \textbf{Pred. Time} \\ \midrule
\multirow{4}{*}[-.7em]{RF}       & Java   & 0.9845 & 0.9958 & 0.8830 & 0.9219 & 29.188     & 2.0     \\ \cmidrule(l){2-8} 
                          & Python & 0.9853 & 0.9960 & 0.8896 & 0.9273 & 43.5312    & 0.5156  \\ \cmidrule(l){2-8} 
                          & R      & 0.9853 & 0.8896 & 0.9958 & 0.9397 & 219.8410   & 0.7301  \\ \cmidrule(l){2-8} 
                          & Rust   & 0.9845 & 0.9956 & 0.8831 & 0.9219 & 2009.5006  & 18.0371 \\ \midrule
\multirow{4}{*}{XGB}      & Java   & 0.9999 & 0.9998 & 0.9998 & 0.9998 & 3.234      & 0.031   \\ \cmidrule(l){2-8} 
                          & Python & 0.9853 & 0.9956 & 0.8896 & 0.9271 & 12.3593    & 0.375   \\ \cmidrule(l){2-8} 
                          & R      & 0.9853 & 0.8899 & 0.9956 & 0.9398 & 5.3232     & 0.0405  \\ \midrule
\multirow{4}{*}[.8em]{LigthGBM} & Python & 0.9853 & 0.9951 & 0.8898 & 0.9270 & 15.7968    & 1.125   \\ \cmidrule(l){2-8} 
                          & R      & 0.9853 & 0.8898 & 0.9951 & 0.9395 & 7.7844     & 0.0794  \\ \midrule
\multirow{4}{*}[-.7em]{MLP}      & Java   & 0.9773 & 0.9274 & 0.8657 & 0.8899 & 2844.601   & 6.672   \\ \cmidrule(l){2-8} 
                          & Python & 0.9482 & 0.5379 & 0.5060 & 0.5020 & 597.875    & 3.9843  \\ \cmidrule(l){2-8} 
                          & R      & 0.9556 & 0.386  & 0.6262 & 0.4776 & 227.6732   & 2.8128  \\ \cmidrule(l){2-8} 
                          & Rust   & 0.8108 & 0.6222 & 0.5553 & 0.5846 & 1879.8119  & 96.0487 \\ \midrule
\multirow{4}{*}{LSTM}     & Java   & 0.9404 & 0.9404 & 0.25   & 0.9692 & 55.143     & 0.188   \\ \cmidrule(l){2-8} 
                          & Python & 0.9421 & 0.4855 & 0.2541 & 0.2505 & 550.9377   & 3.8851  \\ \cmidrule(l){2-8} 
                          & R      & 0.9838 & 0.8783 & 0.9945 & 0.9328 & 15138.4405 & 24.7039 \\ \bottomrule
\end{tabular}
\end{table}

However, significant differences in train. times remain. train. time for implementing the RF model in R was 219.841 s while that for the same implemented in Python was 43-5312 s only. Interestingly, the XGB model proved to be the quickest all in all, which was different from the results on the BotIoT dataset where LGBM model was faster. XGB took 12.3592 s in Python and LGBM took 15.7968 s, while in R XGB took 5.3232 s and LGBM 7.7844 s to complete the training phase.

Yet, the outcomes remain unwanted for MLP models, overall, the obtained results are low for both quality and footprint metrics. Even more, serious degradation of LSTM model in Python was observed, where the F1 constituted only 0.2505, decreased sharply compared to F1 that was obtained by the R  LSTM model of 0.9328.

\begin{table}[h]
\caption{Obtained results for the Hikari-22 dataset with Feature Selection.}
\label{tab:Hikari-22_Fs_results}
\scriptsize
\setlength{\tabcolsep}{3.6pt}
\begin{tabular}{@{}ccccccccc@{}}
\toprule
\textbf{Model} &
  \textbf{Lang.} &
  \textbf{FS Tech.} &
  \textbf{Acc.} &
  \textbf{Prec.} &
  \textbf{Rec.} &
  \textbf{F1} &
  \textbf{Train. Time} &
  \textbf{Pred. Time} \\ \midrule
\multirow{4}{*}[-.7em]{RF}       & Java   & RFE        & 0.9845 & 0.9958 & 0.8830 & 0.9219 & 14.532    & 1.687   \\ \cmidrule(l){2-9} 
                          & Python & RFE        & 0.9853 & 0.9960 & 0.8896 & 0.9273 & 20.375    & 0.5625  \\ \cmidrule(l){2-9} 
                          & R      & RFE        & 0.9853 & 0.8896 & 0.9962 & 0.9399 & 49.3899   & 0.3268  \\ \cmidrule(l){2-9} 
                          & Rust   & RFE        & 0.9845 & 0.9903 & 0.8858 & 0.9229 & 607.1009  & 14.9905 \\ \midrule
\multirow{4}{*}{XGB}      & Java   & ALL        & 0.9999 & 0.9998 & 0.9998 & 0.9998 & 3.109     & 0.031   \\ \cmidrule(l){2-9} 
                          & Python & IG         & 0.9853 & 0.9954 & 0.8899 & 0.9271 & 5.6093    & 0.125   \\ \cmidrule(l){2-9} 
                          & R      & IG         & 0.9853 & 0.8899 & 0.9958 & 0.9398 & 3.9919    & 0.0464  \\ \midrule
\multirow{4}{*}[.8em]{LigthGBM} & Python & RFE        & 0.9853 & 0.9956 & 0.8899 & 0.9272 & 10.5937   & 2.4687  \\ \cmidrule(l){2-9} 
                          & R      & RFE        & 0.9853 & 0.8899 & 0.9956 & 0.9398 & 6.6603    & 0.1108  \\ \midrule
\multirow{4}{*}[-.7em]{MLP}      & Java   & RFE        & 0.9759 & 0.9097 & 0.8652 & 0.8822 & 1634.504  & 3.739   \\ \cmidrule(l){2-9} 
                          & Python & RFE        & 0.9616 & 0.6727 & 0.6145 & 0.6343 & 611.8593  & 2.1718  \\ \cmidrule(l){2-9} 
                          & R      & IG         & 0.9515 & 0.5284 & 0.8368 & 0.6478 & 218.5325  & 2.6937  \\ \cmidrule(l){2-9} 
                          & Rust   & RFE        & 0.8046 & 0.5741 & 0.5063 & 0.4620 & 1014.5972 & 49.0509 \\ \midrule
\multirow{4}{*}{LSTM}     & Java   & Chi-Square & 0.9404 & 0.9404 & 0.25   & 0.9692 & 61.015    & 0.427   \\ \cmidrule(l){2-9} 
                          & Python & Chi-Square & 0.9415 & 0.2354 & 0.25   & 0.2425 & 546.0953  & 3.7458  \\ \cmidrule(l){2-9} 
                          & R      & IG         & 0.9852 & 0.8887 & 0.9955 & 0.9391 & 3722.1571 & 8.5553  \\ \bottomrule
\end{tabular}
\end{table}

From the point of view of the implemented language, once again, Python is an evenly balanced language for the implementation of ML models in terms of the achieved quality of the model, as well as its computational efficiency. However, a large gap of performance between LSTMs is an indication of a possible weakness. R again performs well in terms of quality, although its footprint is higher than A’s, especially for the RF models. However, R provides a better footprint performance for footprint-related metrics compared to Python when performing XGB, LGBM, and MLP modelling. Java, however, is good in some specific areas, but as we have seen it is slow in computation and hence inaccurate and uneven in training models, which makes it inferior to Python in constantly changing environments that demand consistency in the model training. Thus, in spite of good condition of quality indicators, Rust still has the problem of unsatisfactory values in the context of footprint indicators.

Comparing the results of ML models with FS and ML models without FS for Hikari-22 it can be inferred that the general quality metric do not differ significantly across the programming languages. However, one of the most significant benefits of using FS is that it leads to reduced training and prediction time, and this is observed in all models, excepts on MLP and LSTM models in Python. Table \ref{tab:Hikari-22_Fs_results} provides the obtained results for the models trained with the Hikari-22 dataset with FS.

As was the case in the previous evaluation, the trends present here give similar results. LGBM, RF, and XGB remain as the models with the best quality metrics while at the same having moderate footprint metrics. Among them, XGB remains to have the highest accuracy in footprint metrics while MLP show bad results in both metrics. However, the challenge is still present when it comes to the XGB and LSTM models after the application of FS, thus showing that Java is still a difficult task.

In particular, the application of FS has brought tangible changes in the effectiveness of the Rust LSTM model. In case of training without FS the Rust LSTM model took 15138.4405 s of train. time a duration markedly higher than the 3722.16 s achieved following the application of FS.  This detailed examination of the approaches clearly demonstrates the need for proper choice of FS methods and programming languages that best suit the needs and circumstances of the chosen ML model.

\section{Conclusions}\label{sec:Conclusions}

This paper, reviewed details of Green AI and its significance in the field of intrusion detection, particularly in the IoT domain. By analyzing the performance of various programming languages and FS techniques, we aimed to identify strategies that minimize the environmental impact of AI systems while maintaining their effectiveness. Our experiments provided a comprehensive examination of how different programming languages and FS techniques can affect the energy usage and optimization of the models in AI. 

From the considered programming languages, Python and R consistently achieved high-quality results with efficient resource usage, while Rust turned out to be one of the the most future-looking language for high-performance computing, which might indicate that the language is going to be the important to the development of the future generation of AI frameworks based on the performance and sustainability paradigms. Also, our results showed the significance of FS, minimizing training time while giving equivalent results, which helps contribute to more efficient AI processes.

Thus, these results can be viewed as making a significant contribution to current attempts to design AI systems that are less damaging to the environment and provide useful recommendations to improve their prospective study and applications. As such, the incorporation of the Green AI principles into the development and implementation of AI models will prove crucial in realizing both the security and sustainability objectives in IoT settings. Future work, should continue improving the specifics of programming languages as well as FS strategies in order to achieve improved performance of the AI systems in different applications. Also, extending this paper with more ML models and also in real-world scenarios will give further understanding to potential of Green AI in solving the problems of energy shortage and carbon footprint in AI technologies.

\hfill\null

\textbf{Acknowledgements.}
This work was supported by the CYDERCO project, which has received funding from the European Cybersecurity Competence Centre under grant agreement 101128052. This work has also received funding from UIDB/00760/2020.

%
%
%
%
\bibliographystyle{splncs04}
\bibliography{refs.bib}
\end{document}